# Malaria Incidence in the Philippines: Prediction using the Autoregressive Moving Average Models

Empha Grace Perez[1] and Roel F. Ceballos[2]

Department of Mathematics and Statistics
University of Southeastern Philippines
Bo. Obrero, Davao City, Philippines 8000
[1]emphagrace_perez@yahoo.com
[2]roel.ceballos@usep.edu.ph

## ABSTRACT

*The study was conducted to develop an appropriate model that could predict the weekly reported Malaria incidence in the Philippines using the Box-Jenkins method. The data were retrieved from the Department of Health(DOH) website in the Philippines. It contains 70 data points of which 60 data points were used in model building and the remaining 10 data points were used for forecast evaluation. The R Statistical Software was used to do all the necessary computations in the study. Box-Cox Transformation and Differencing was done to make the series stationary. Based on the results of the analysis, ARIMA (2, 1, 0) is the appropriate model for the weekly Malaria incidence in the Philippines.*

## 1 Introduction

Malaria is one of the worlds deadliest diseases. It is caused by a Plasmodium parasite and transferred to human by means of the bite of Anopheles mosquito (Lam, 2017). Every year, more than one million people die from Malaria and most victims were children under five years of age. According to the United Nations Childrens Fund(UNICEF), approximately 1,200 children die everyday or fifty(50) children die every hour because of Malaria. In 2015, the World Health Organization(WHO) reported that there were 212 million cases of Malaria and 429,000 estimated deaths worldwide. Approximately, three hundred thirty thousand (330,000) of these deaths were children under five years of age.

In recent years, Malaria has been eliminated from many developed countries with temperate climates. However, the disease remains a major health problem in many developing countries, in tropical and subtropical parts of the world(Centers for Disease Control and Prevention, 2017). Due to severe health impact of Malaria epidemics, there is a growing need for methods that will allow accurate forecasting of future incidences. Over the past years, many new statistical models have been developed. Appiah, Otto and Nabubie(2015) have found out that

ARIMA (2,1,1) can predict the Malaria cases in Ejisu-Juaben Municipality in Ghana Africa.
In the Philippines, programs and policies on health monitoring has been created by the Department of Health to eradicate high incidence of diseases such as Malaria. Several studies have been conducted to find ways to improve the existing health care services provided by the Philippine government. Carillo, Largo and Ceballos (2018) conducted a principal component analysis on the Philippine health data to determine the underlying structures of its different determininants.Results show that importance of safe water supply and emphasis on child and women's health and importance of the Barangay Health Workers and stations are the components that can summarized the Philippine Health data.

This study aims to investigate the characteristics of the weekly reported cases of Malaria incidence and to find the best model for prediction. R is a programming language that is primarily used for statistical computing and graphics will be used to do all the necessary calculations. It was created by Ross Ihaka and Robert Gentleman at the University of Auckland, New Zealand, and is currently developed by the R Development Core Team. Moreover, it provides a wide variety of statistical and graphical techniques, and is highly extensible. R software was used in producing plots and computations in this study. The R software includes packages such as: tseries for testing stationarity;fpp for model estimation, model diagnostics, accuracy measures, graphical presentations, and for forecasting procedure; and astsa for getting the numerical values of the ACF and PACF of the time series.

## 2 Methods

According to Montgomery, Jennings and Kulahci (2008), Box-Jenkins forecasting method consists of a three-step iterative procedure as follows: Model Identification, Model Estimation and Diagnostic Checking. An additional step called Model Evaluation is also suggested. Thus, the resulting procedure is as follows:

### 2.1 Model Identification

The first thing to consider in forecasting is to determine if the series is stationary by checking if the mean and variance are stable. Two approaches can be used to check the stationarity of the series. These are ACF plots and Augmented Dickey Fuller (ADF) test. If the series is not stationary, transformation using one or the combination of the following techniques can be done in order to achieve stationarity.

1. We can difference the time series data. That is, given the series $Y_t$, we have,

$$Z_t = Y_t - Y_{t-1}$$

2. We can transform the data using Box-Cox transformation with the appropriate $\lambda$ values in order to stabilize the variance of the series or when the residuals of the model may exhibit some problems.

Once the data is stationary, candidate models will be identified using the ACF and PACF plots. The final model will be chosen from the candidate models using the smallest value of the Akaike Information Criterion(AIC).

## 2.2 Model Estimation

Once the final model is identified, the next thing to consider for to estimate the parameters of the model. The estimation of parameters will be done using maximum likelihood estimation (MLE). That is, the values of the parameters of the orders $p$ and $q$ as well as the constant term and the residuals are obtained.

## 2.3 Diagnostic Checking

After the estimation of the model parameters, residual analysis was done to how well the model fits the data. The model is considered to have a good fit if there is no visible pattern in the residual plots, that is, there is no autocorrelation in the residuals of the model. The Ljung-Box Test will be used as a formal test to determine if the model is a good fit to the series.If all the results implied a lack of fit of the model to the series, then go back to step 2 and try to develop a better model.

## 2.4 Model Evaluation

Model evaluation is used to evaluate forecast accuracy of the model. The forecast errors are obtained by taking the difference between the 10 data points that were not part of the model building (actual values) and the one-step ahead forecasts. It is ideal that the forecast errors behave like a Gaussian white noise. If there are no significant spikes in the ACF and PACF plots of the forecast errors then it is considered a white noise. The Shapiro-Wilk test will be used to test for the normality of the forecast errors.

## 3 Statistical Treatment

1. Autocorrelation Function (ACF). ACF Plot can be used to determine the stationarity of the series. The ACF plot of a stationary series will drop quicky to zero while for the non-stationary series is slowly decaying. Also, ACF is useful in identifying the order of a Moving Average Model. Given a time series $Y_t$, the sample autocorrelation function at lag k is

$$r_k = \frac{E[(Y_t - \mu)(Y_{t+k} - \mu)]}{E[(Y_t - \mu)^2]}$$

   where $E$ is the expected value operator.

2. Partial Autocorrelation Function (PACF). PACF is useful in identifying the order of an Autoregressive Model. Given a time series $Y_t$, the partial autocorrelation of lag k, is the autocorrelation between $Y_t$ and $Y_{t+k}$ is

$$\alpha(1) = Cor(Y_{t+1}, Y_t) \text{ for } k = 1$$
$$\alpha(k) = Cor[Y_{t+k} - P_{t,k}(Y_{t+k}), Y_t - P_{t,k}(Y_t)] \text{ for } k \geq 2$$

where $P_{t,k}(x)$ denotes the projection of $X$ onto the space spanned by $X_t + 1, ..., X_{t+k-1}$.

3. Augmented Dickey Fuller (ADF) Test. ADF Test is a unit root test for stationarity. The null hypothesis for this test is there is a unit root and the alternative hypothesis for this test is the data is stationary. The ADF test statistic is

$$DF = \frac{\gamma}{SE(\gamma)}$$

where $\gamma$ is the least square estimate and $SE(\gamma)$ is the standard error.

4. Akaike's Information Criterion (AIC). AIC is used to judge a model by how closed its fitted values in terms of certain expected values. The criterion value assigned to a model is only meant to rank competing models and tell the best among the given alternatives. In general, AIC is calculated using the relation

$$AIC = 2k - 2ln(L)$$

where $k$ is the number of parameters in the model and $L$ is the maximized value of the likelihood function.

5. Ljung-Box Test. The Ljung-Box test is a diagnostic tool used to test the lack of fit of a time series model. The null hypothesis of the test states that the model does not exhibit lack of fit and the alternative hypothesis states that the model exhibits lack of fit. Given a time series Y of length n, the test statistic is

$$Q_K = n(n+2) \sum_{i=1}^{m} \frac{r_k^2}{n-k}$$

where $r_k$ is the autocorrelation of the series at lag $k$ and $m$ is the number of lags being tested.

6. Box-Cox Transformation. If the variance of the series is non-constant or if the residuals of the model exhibit lack of fit, Box-Cox Transformation is used. An exponent lambda ($\lambda$) is at the core of the Box-Cox Transformation. The optimal value of $\lambda$ will be selected and used. Given a time series $Y_t$ the formula is

$$W_t = \frac{Y_t^\lambda - 1}{\lambda}$$

if $\lambda$ is not equal to zero. If $\lambda = 0$, then we have,

$$W_t = log Y_t.$$

7. Shapiro-Wilk Test. The Shapiro-Wilk test is way to tell if a random sample comes from a normal distribution. The null hypothesis for this test is that the data comes from a normal distribution and the alternative hypothesis for this test is that the data does not come from a normal distribution. The test gives a $W$ value. Small W is evidence of departure from normality. The test statistic of the test is

$$W = (\Sigma_{i=1}^{n} a_i x_i)^2 \; \Sigma_{i=1}^{n} (x_{i_1} - x)^2 -$$

where $x_i$ are the ordered sample values and $a_i$ are constants generated from the covariances, variances and of order statistics of a sample size $n$ from a normal distribution.

## 4 Results and Discussion

Figure 1 displays the time series plot($Y_t$) of the reported weekly Malaria incidence in the Philippines. The plot shows 60 weekly observations from January 1, 2016 up to February 25, 2017. The maximum reported Malaria incidence is 491 which can be observed on week 24 (June 12 - 18, 2016) and the minimum is 1 which can be observed on week 5 (January 31, 2016 - February 6, 2016). This implies that Malaria incidence in the Philippines ranges from 1 to 491 on a weekly basis. In addition, the trend of the data is increasing from week 1 up to week 25 and decreasing onwards. Also, the figure shows no clear seasonality in the data and it is hard to conclude whether the time series is stationary or not.

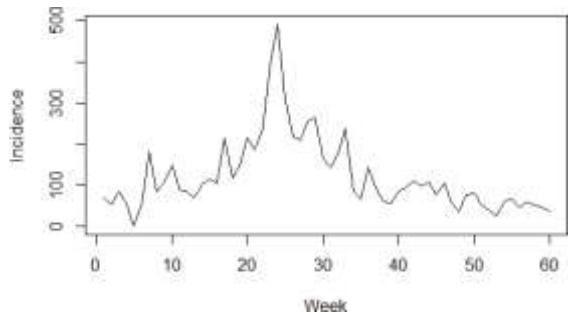

Figure 1: Weekly Malaria incidence in the Philippines ($Y_t$)

To formally test the stationarity of the series, the ADF test for stationarity was used. Table 1 shows the ADF test statistic value and the p-value. Since the pvalue is large, the null hypothesis is not rejected. Hence, the series has a unit root and transformation is needed to make the $Y_t$ stationary. The first difference was obtained ($W_t = Y_t - Y_{t-1}$) inorder to make the series stationairy. However, despite achieving stationarity after differencing (see Table 2) the transformation was insufficient since one of the coefficients of the chosen model was not significant (see Table 3) and the residuals of the the chosen model is not normal (see Table 4).

Thus, Box-Cox Transformation is done prior to differencing to find a better model. Figure 2 shows the time series plot of the tansformed and differenced series ($Z_t$). It can be observed from the plots of $Z_t$ that the series seems to be stationary. It can be also observed that the time series data has no apparent trend or seasonality.

Table 1: ADF Test Results for $Y_t$

| ADF Test Statistic | p-value |
|---|---|
| -1.23 | 0.602 |

Table 2: ADF Test Results for $W_t$

| ADF Test Statistic | p-value |
|---|---|
| -4.63 | 0.01 |

To test the stationarity of the series, ADF test was applied to $Z_t$. Table 5 shows the test statistic value and its p-value. Since the p-value is 0.01, the null hypothesis is rejected. Hence, the series has no unit root which implies that the series is now stationary. Figure 3 shows the ACF and PACF plots of the differenced series ($Z_t$). It can be observed that both autocorrelation and partial autocorrelation are significant at lag 2.

Based on the ACF and PACF plots, the following ARIMA models are considered: ARIMA (0, 1, 2), ARIMA (2, 1, 0) and ARIMA (2, 1, 2). Table 6 shows the tentative models for the weekly Malaria incidence ($Z_t$). It also shows the Akaike Information Criterion (AIC) for each tentative model. The AIC is used to evaluate a model. The model with the least AIC value shall be selected. Among the tentative models, ARIMA (2, 1, 0) has the least AIC value. Thus, it is the chosen model for the weekly Malaria incidence.

Table 7 shows the estimated coefficients, standard error, z-value and the p-value of the ARIMA(2,1,0). It can be observed that the p-value of AR(1) and AR(2) are less than 0.05 level of significance. Thus, the estimates of the autoregressive parameters are significantly different from zero.

## 4.1 Residual Analysis and Forecast Evaluation

Figure 4 and 5 shows the residual versus time and sample quantile versus theoretical quantile plots. These plots are used to assess how well the chosen model fits the data. From the given plots, it can be observed in Figure 4 that the residuals fluctuate randomly around 0. Hence, it can be concluded that there is no autocorrelation among the residuals and that the plot shows no visible pattern. Thus, the variance is stable. Also, from Figure 5 (Sample Quantiles Vs Theoretical Quantiles), it can be observed that the residuals are near the theoretical line. Therefore, the residuals are normally distributed. Also, the ACF and PACF plots of residuals in

Table 3: Model Parameters for $W_t$

| Model Statistics | AR(1) | AR(2) |
|---|---|---|
| Estimate | -0.0889 | -0..4198 |
| Standard Error | 0.1171 | 0.1151 |
| z-value | -0.7592 | -3.6473 |
| p-value | 0.4477 | 0.01 |

Table 4: Test for Normality of Residuals for $W_t$

| Shapiro-Wilk Test Statistic | p-value |
|---|---|
| 0.9597 | 0.04 |

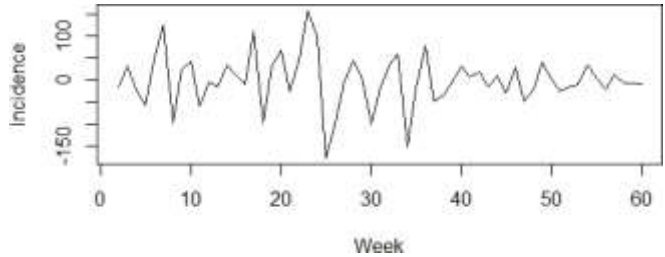

Figure 2: Time series plot of $Z_t$

Table 5: ADF Test Results $Z_t$

| ADF Test Statistic | p-value |
|---|---|
| -6.12 | 0.01 |

Table 6: Tentative Models for $Z_t$

| Tentative Models | AIC |
|---|---|
| ARIMA(0,1,2) | 298.99 |
| ARIMA(2,1,0) | 296.80 |
| ARIMA(2,1,2) | 299.07 |

Table 7: Tentative Models for $Z_t$

| Model Statistics | AR(1) | AR(2) |
|---|---|---|
| Estimate | -0.22712 | -0.50015 |
| Standard Error | 0.11156 | 0.10949 |
| z-value | -2.0359 | -4.5681 |
| p-value | 0.0418 | 0.01 |

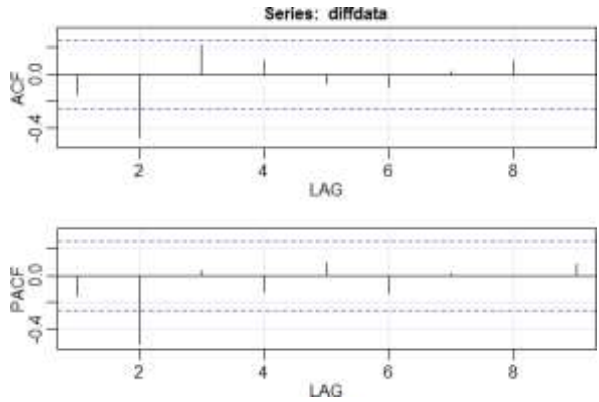

Figure 3: AFC and PACF Plots of $Z_t$

Figure 6 show that individual values of the ACF and PACF are within acceptable limits. Hence, they are not significantly different from zero.

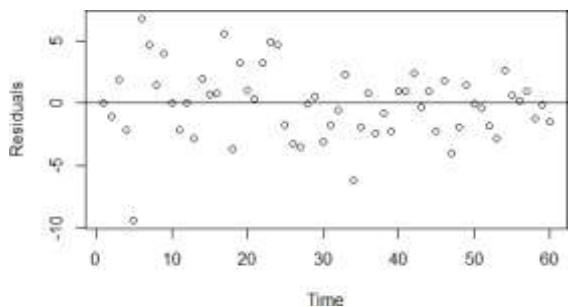

Figure 4: Plot of Residuals Vs Time

To formally check if the residual of the chosen model is uncorrelated, Ljung-Box test was applied. Table 8 shows the Ljung-Box Test value and p-value. Since the p-value is greater than 0.05, there is not enough evidence to say that autocorrelation exist among the residuals. Thus, the chosen model is appropriate for the series.

Table 8: Ljung-Box Test

| Ljung-Box Q* Statistic | p-value |
|---|---|
| 4.1824 | 0.8403 |

Table 9 shows the one-step ahead forecast values, actual values and forecast errors of ARIMA (2, 1, 0).

Figure 7 shows the ACF and PACF plots of the forecast errors respectively. Since the autocorrelation and partial autocorrelation values of the forecast errors are within the limits, the values are not significantly different from zero. Therefore, the forecast errors are considered a white noise. To formally check if the forecast errors are normally distributed, Shapiro-Wilk test statistic and the p-value of the test is presented in Table 10. Since the p-value is large, the null hypothesis cannot be rejected. Thus the forecast errors are normally distributed. The forecast errors are considered Gaussian white noise.

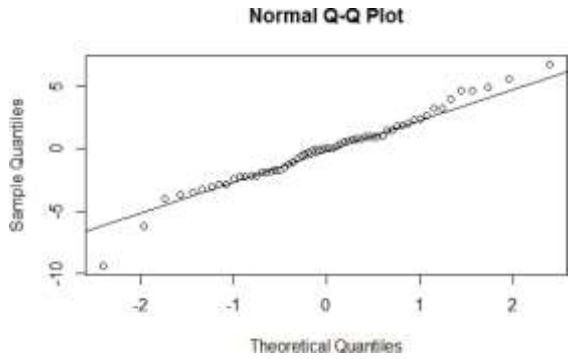

Figure 5: Plot of Sample and Theoretical Quantiles

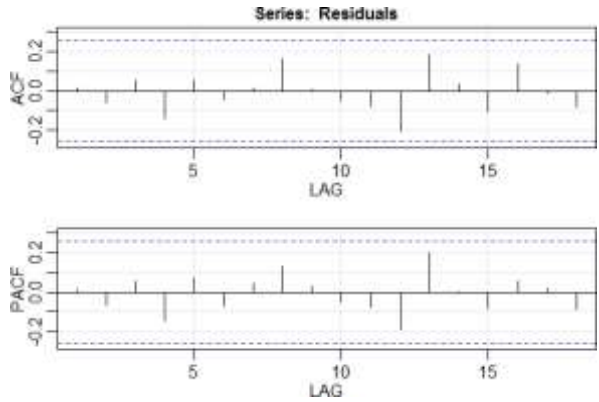

Figure 6: ACF and PACF of Residuals

Table 9: One-step ahead Forecast

| Week | Date | Actual Values | One-step ahead Forecast | Forecast Errors |
| --- | --- | --- | --- | --- |
| Week 9 | 26 Feb- 04 Mar | 13 | 28 | -15 |
| Week 10 | 05 Mar-11 Mar | 3 | 16 | -13 |
| Week 11 | 12 Mar-08 Mar | 2 | 9 | -7 |
| Week 12 | 19 Mar- 25 Mar | 16 | 6 | 10 |
| Week 13 | 26 Mar - 01 Apr | 6 | 11 | -5 |
| Week 14 | 02 Apr - 08 Apr | 16 | 3 | 13 |
| Week 15 | 09 Apr - 15 Apr | 8 | 16 | -8 |
| Week 16 | 16 Ap - 22 Apr | 19 | 5 | 14 |
| Week 17 | 23 Apr - 29 Apr | 20 | 16 | 4 |
| Week 18 | 30 Apr - 06 May | 2 | 11 | -9 |

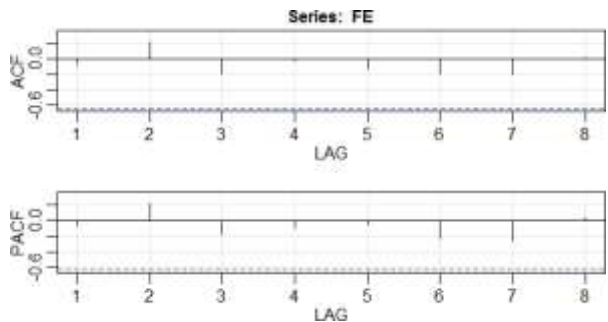

Figure 7: ACF and PACF of Forecast Errors

Table 10: Test for Normality of Forecast Errors

| Shapiro-Wilk Test Statistic | p-value |
|---|---|
| 0.88724 | 0.1578 |

The final model for weekly reported Malaria incidence in the Philippines is ARIMA (2, 1, 0). The diagnostic checks yield satisfactory results. Thus, the ARIMA(2,1,0) is an appropriate model to predict the weekly Malaria incidence in the Philippines.